\documentclass[12pt]{article}
\usepackage{epsf}
\usepackage{aas_macros}
\usepackage{amssymb}
\usepackage{amsmath}
\usepackage{braket}
\usepackage{mathrsfs}
\usepackage{mathtools}
\usepackage{array}
\usepackage{fancyhdr}
\usepackage[dvipdfmx]{graphicx}
\usepackage{color}
\usepackage{cite}
\usepackage{bm}
\usepackage{url}
\usepackage{fancyhdr}
\usepackage[colorlinks,citecolor=blue]{hyperref}
\usepackage[hang,small,bf]{caption}
\usepackage[subrefformat=parens]{subcaption}
\renewcommand{\thefootnote}{\#\arabic{footnote}}

\renewcommand{\thefootnote}{\fnsymbol{footnote}}
\def\thefootnote{\fnsymbol{footnote}}

\makeatletter

\@addtoreset{equation}{section}
\makeatother

\def\be{\begin{equation}}
\def\ee{\end{equation}}
\def\ben{\begin{eqnarray}}
\def\een{\end{eqnarray}}

\begin{document}

%\begin{titlepage}

\begin{center}

\vskip .75in

{\Large \bf On the sound velocity bound in neutron stars}

\vskip .75in

{\large
Shrijan Roy$\,^1$,  Teruaki Suyama$\,^2$
}

\vskip 0.25in

{\em
$^{1}$Department of Physics, Indian Institute of Technology Delhi, Hauz Khas, New Delhi 110016, India\\
$^{2}$Department of Physics, Tokyo Institute of Technology, 2-12-1 Ookayama, Meguro-ku,
Tokyo 152-8551, Japan
}

\end{center}
\vskip .5in

\begin{abstract}
It has been suggested in the literature that the sound velocity
of the nuclear matter $v_s$ violates the so-called sound velocity bound
$v_s \le c/\sqrt{3}$ at high density,
where $c$ is the speed of light.
In this paper, we revisit this issue and confront the current measurements of  
mass, radius, and tidal deformability of neutron stars with 
$10^5$ different equations of state which are parametrized at low density 
and saturates the sound velocity bound beyond twice the saturation  
density where the equation of state has not been constrained yet,
by which we can conservatively obtain the maximum mass of the neutron stars 
compatible both with the observed properties of neutron stars and
the sound velocity bound.
We find that majority of the models are eliminated by the incompatibility with the observations and, especially, the recently detected massive pulsar ($2.35\pm 0.17 M_\odot$) is hardly realized by our simulations.
Our study strongly supports the violation of the sound velocity bound.
\end{abstract}

%\end{titlepage}

\renewcommand{\thepage}{\arabic{page}}
\setcounter{page}{1}
\renewcommand{\thefootnote}{\#\arabic{footnote}}
\setcounter{footnote}{0}

\section{Introduction}
Even though the quantum chromodynamics (QCD) has long been established as a fundamental theory of strong interaction, 
properties of nuclear matter beyond the saturation density are poorly understood both theoretically and experimentally. 
Since neutron stars (NSs) are very compact stellar objects consisting of nuclear matter 
and the matter density inside can exceed the saturation density, 
observed properties of NSs such as mass, radius, and tidal deformability enable us to constrain the 
equation of state (EoS) of nuclear matter at such large density.
At zero temperature, which is a good approximation for NSs, EoS refers to the pressure
as a function of the energy density: $P=P(\epsilon)$.

Observations of massive NSs exceeding 2~$M_\odot$ \cite{Romani:2022jhd} 
suggest that pressure rapidly increases as the energy density increases
to support the star against gravity. 
The slope of the EoS is related to the sound velocity of the nuclear matter as $v_s^2=c^2\frac{\partial P}{\partial \epsilon}$. 
In this respect, there has been a growing interest in whether the sound velocity $v_s^2$ 
exceeds the so-called conformal limit $\frac{c^2}{3}$ at some density or not (i.e., \cite{Kojo:2020krb}). 
Obviously, the bound $v_s^2 \le \frac{1}{3}c^2$ is satisfied for systems consisting of non-interacting particles
or interacting non-relativistic particles.
At ultra-density where interactions between ultra-relativistic quarks become negligible 
due to asymptotic freedom, 
$v_s \to \frac{c}{\sqrt{3}}$ and the bound is also satisfied there. 
It is known that the bound is satisfied for a wide class of field theories (e.g., \cite{Cherman:2009tw}).
Meanwhile, recent studies show that the bound can be largely 
violated in theories close to QCD \cite{Riley:2021pdl, McLerran:2018hbz, Leonhardt:2019fua, Margueron:2021dtx}
although it is still not theoretically established that the bound is violated in QCD.
Thus, there is a strong motivation to investigate if the bound is satisfied or violated by 
studying NSs.

By making comparison between various possible EoSs and observations of massive pulsars 
and measurements of tidal deformability of NSs by gravitational-wave experiments,
recent papers strongly indicate that the bound is actually violated and the sound velocity takes
a local maximum (larger than $\frac{1}{\sqrt{3}}c$) at intermediate nuclear density \cite{Bedaque:2014sqa, Altiparmak:2022bke, Ecker:2022xxj, Brandes:2022nxa, Braun:2022olp, Ecker:2022dlg}.
This has been done by adopting the known EoS at low density and connecting it to higher density regime where
EoS is given by piecewise form (linear in the chemical potential or in the energy density) or some specific functional form characterized with variable parameters. 
In particular, the recent study \cite{Ecker:2022dlg} incorporating the heaviest NS ($M=2.35 \pm 0.17~M_\odot$) 
recently discovered as a pulsar shows that the bound is violated even if the lowest 
value in the allowed mass range (i.e. $M=2.18~M_\odot$) is adopted.

Although the existing studies already provide a strong support in favor of the violation of the bound,
we would like to address this issue by adopting the type of EoS used in \cite{Bedaque:2014sqa}.
This EoS matches the EoS of nuclear matter below twice of saturation density where
the EoS is thought to be well understood. 
Above the critical density, the EoS sharply turns into stiff one given by $v_s=\frac{1}{\sqrt{3}}c$ at any density.
Obviously, this EoS is artificial and unrealistic. 
Yet, the benefit from using it is that we can obtain
the robust maximum mass of the NS compatible with the bound.
In other words, observation of a NS heavier than this maximum mass provides an absolute evidence
that the sound velocity exceeds $\frac{1}{\sqrt{3}}c$ irrespective of the uncertainties of EoS at large density.
Notice that our approach is the same philosophy as the one used to obtain the robust maximum mass of NS consistent with causality 
$v_s \le c$ \cite{Rhoades:1974fn} by adopting the EoS which becomes the stiffest 
one given by $v_s =c$ above a certain density .

In the next section, we first define the EoS used in this paper based on some relevant papers.
We then briefly review how to compute the tidal deformability and the effect of rotation on NS masses.
In Sec.~\ref{ReDi}, we present our results, followed by discussions about the implications and comparison with
previous studies.
Final section is devoted to the conclusion.

\section{Neutron Star Structure}
\label{NSS}
\subsection{$\beta$-equilibrated matter}
We use the Skyrme based parametrization~\cite{SKYRME1958615} refined by Hebeler et. al~\cite{Hebeler_2013} under which the energy density including the major contributions from rest masses(electron mass neglected) can be written as 
\begin{align}
\frac{\epsilon}{n_0T_0} =& \frac{3 \overline{n}^{5/3}}{5} \left[x^{5/3} + (1-x)^{5/3}\right]2^{2/3} - \overline{n}^2\left[(2\alpha - 4\alpha_L)x(1-x) + \alpha_L\right] \nonumber
\\ &+ \overline{n}^{\gamma+1}\left[(2\eta - 4\eta_L)x(1-x) + \eta_L\right] + \overline{n}\frac{\left[(1-x)M_n + xM_p\right]}{T_0}c^2 \text{ ,}
\label{e-d}
\end{align}
where $x = n_p/n$ represents the proton fraction present, $n_0$ = 0.16 $\pm$ 0.01 fm$^{-3}$, $T_0$ = $\left(\frac{3 \pi^2 n_0}{2}\right)^{2/3} \hbar^2/2m$. The set of parameters $\{\alpha,\eta,\alpha_L,\eta_L,\gamma\}$, control how the EoS will act and vary as a function of $\overline{n} = n/n_0$. The pressure corresponding to Eq.~(\ref{e-d}) is given by
$p = n^2\frac{\partial (\epsilon/n)}{\partial n}$. 
Due to the symmetric nature of matter at saturation, we have
\begin{equation}
p_{\overline{n}=n_0,x=1/2} = 0.
\label{pres_zero}
\end{equation}
Also one may define the quantity $B$ such that, 
\begin{equation}
-B = \frac{\epsilon(\overline{n}=n_0,x=1/2)}{n_0} - \left(\frac{M_n + M_p}{2}\right) \text{ ,}
\label{B_0}
\end{equation}
which is also computed at the saturation density as 16 $\pm$ 0.1 MeV ~\cite{PhysRevC.89.054314}. Additionally the EoS can be described in terms of some properties : (a) The nuclear incompressibility coefficient $K$ which measures the stiffness of symmetric nuclear matter can be described as
$K(n) = 9\frac{\partial p}{\partial n}$.
At saturation density in the symmetric phase, it becomes
\begin{equation}
K_0 = 9n_0^2\frac{\partial^2(\epsilon/n)}{\partial n^2}\bigg\rvert_{\overline{n}=n_0,x=1/2} \text{ .}
\label{K0}
\end{equation}
From study of giant resonances~\cite{PhysRevC.69.034315,Shlomo2006}, $K_0$ has been determined to be 235 $\pm$ 25 MeV. (b) The symmetry energy $S(n)$ is defined by
\begin{equation}
S(n) = \frac{1}{8}\frac{\partial^2 (\epsilon/n)}{\partial n^2} \text{ .}
\label{S0}
\end{equation}
Additionally we may also define the slope of this symmetric energy curve at saturation density by 
\begin{equation}
L(n_0) = \frac{3n_0}{8}\frac{\partial^3 (\epsilon/n)}{\partial n \partial x^2} \text{ .}
\label{L0}
\end{equation}
Using experimental data from heavy
ion collisions, giant dipole resonances, and dipole polarizabilities~\cite{Lattimer_2013,Lattimer2014}, the values of $S$ and $L$ were taken to be 32 $\pm$ 2 MeV and 50 $\pm$ 15 MeV.

The set of these previous 5 equations can be used to solve for the values of the parameters $\{\alpha,\eta,\alpha_L,\eta_L,\gamma\}$ characterizing the EoS. 
Furthermore, the proton fraction $x$ can be computed by the condition
for the $\beta$-equillibrium of matter ~\cite{Mondal_2022}:
\begin{equation}
\frac{\partial (\epsilon/n)}{\partial x} + \mu_e = 0.
\label{minim}
\end{equation}
%which can be very intuitively proved by a definition of a chemical potential and using the chemical potential relation $\mu_n = \mu_p + \mu_e$~\cite{Mondal_2022}. 
% We take $\mu_e$ to be $\hbar c(3 \pi^2 x n_0 \overline{n})^{1/3}$ which is true for an ultrarelativistic degenerate electron gas. Speed of sound is intrinsically connected to the structure of the neutron star and EoS by the relation $(v_s/c)^2 = dp/d\epsilon$, and as discussed before, we constrain the square of this value below 1/3. 
Now, the EoS obtained from Eq.~(\ref{e-d}) may be correct at lower densities, 
but fails at higher densities.
As we have discussed in the Introduction, our aim is to determine the maximum mass consistent
not only with the EoS compatible with experimental data at low density but also with
the bound $v_s^2 \le c^2/3$ in a conservative manner. 
Thus we adopt the following form of EoS
similar to the one used in \cite{Bedaque:2014sqa}:
%thus, relevant to our job, in which we would need the limiting cases of masses of neutron star, can be achieved by directly making true the relation $dp/d\epsilon = 1/3$ after a certain cutoff nuclear density, i.e.
\begin{equation}
\begin{split}
\epsilon(\overline{n}) &= \epsilon(\overline{n},x), ~~~~~n < 2n_0 \\
&= \epsilon(\overline{n}=2,x) + 3\left[p(\overline{n},x) - p(\overline{n}=2,x)\right],~~~~~
n > 2n_0, 
\label{epsilon_final}
\end{split}
\end{equation}
where $\epsilon(\overline{n},x)$ is obtained after using $x$ determined
from the minimization condition Eq.~(\ref{minim}).
The latter term in the region above twice the saturation density is added to preserve the continuity of the EoS~\cite{Srinivasan2002}.

In order to incorporate the experimental uncertainties of $n_0, B, K, S, L$ in the determination
of the maximum mass of the neutron star,
we randomly generate the values of these five parameters by treating them
as probabilistic variables obeying 
the normal distribution with each mean and variance given by the central value 
and the error of experiments.
We exclude any sample that are outside the $2\sigma$ range of the normal distribution in order
not to include the outliers which inevitably appear when the sample size becomes large.
In our analysis in the next section, we generate a dataset of 10$^5$ different EoS which are further 
processed according to the observational constraints as we will discuss in \ref{POC}.

\subsection{Mass, radius, and tidal deformability of static neutron stars}

The metric sourced by static and spherically symmetric neutron star can be written in the form
\begin{equation}
ds^2 = -e^{\nu(r)}(cdt)^2 + e^{\Phi(r)}dr^2 + r^2(d\theta^2 + \sin^2\theta d\phi^2).
\end{equation}
Outside the star, $\nu (r)$ and $\Phi (r)$ are given by
\begin{equation}
e^{\nu(r)} = e^{-\Phi(r)} = \left[1 - \frac{2GM}{c^2r}\right],
\label{met_func}
\end{equation}
where $M$ is the mass of the neutron star.

Solving Einstein's field equations and using the energy-momentum tensor for this metric,
we obtain the Tolman-Oppenheimer-Volkoff (TOV) equation~\cite{PhysRev.55.374,doi:10.1073/pnas.20.3.169} \begin{equation}
\frac{dp(r)}{dr} = -\frac{\left(\frac{G \epsilon (r) M(r)}{c^2r^2}\right)\left(1 + \frac{p(r)}{\epsilon(r)}\right)\left(1 + \frac{4 \pi r^3 p(r)}{M(r) c^2}\right)}{\left(1 - \frac{2 G M(r)}{c^2r}\right)},
\label{TOV}
\end{equation}
where the mass $M(r)$ contained in the sphere of radius $r$ is related to the
energy density $\epsilon$ as
\begin{equation}
M(r) = \int_0^r \frac{4 \pi r'^2 \epsilon(r')}{ c^2}dr' \text{ .}
\end{equation}
By solving the TOV equation starting at $r=0$ under a given central pressure
up to the radius $R$ where $p$ drops to zero,
we can obtain the mass and the radius of the neutron star.

The tidal deformability is defined as the ratio of mass quadrupole moment of a star $Q_{ij}$ to the external tidal field $\varepsilon_{ij}$ as~\cite{PhysRevC.95.015801,Hinderer_2008}
\begin{equation}
\lambda = -\frac{Q_{ij}}{\varepsilon_{ij}} = \frac{2k_2R^5}{3G} \text{ ,}
\end{equation}
%the latter, connecting $\lambda$ to the EoS of the Neutron Star, 
where $k_2$ is the Tidal Love number whose analytical formula is given as follows~\cite{Hinderer_2008}:
\begin{align}
k_2 =&  \frac{8C^5}{5}\left((1 - 2C)^2\left[2 + 2C(y-1) - y\right]\right)\{(2C(6 - 3y + 3C(5y - 8)) \nonumber \\ &+4C^3\left[13 - 11y + C(3y-2) + 2C^2(1+y)\right] \nonumber \\
&- 3(1-2C)^2\left[2 - y + 2C(y-1)\right]\ln\left(\frac{1}{1-2C}\right)\}^{-1} \text{ ,}
\end{align}
where $C=GM(R)/c^2R$ and $y=R\beta(R)/H(R)$. 
The latter two unknown quantities can be calculated by solving the set of equations~\cite{Hinderer_2010}:
\begin{equation}
\beta = \frac{dH}{dr} \text{ ,} ~~~~~~
\frac{d\beta}{dr} + X(r)\beta + Y(r)H = 0 \text{ ,}
\end{equation}
where $X(r)$ and $Y(r)$ are represented by
\begin{equation}
X(r) = \left(\frac{2}{r} + \frac{1}{\left(1 - \frac{2 G M(r)}{c^2r}\right)}\left(\frac{2GM(r)}{c^2r^2} + \frac{4\pi r(p - \epsilon)G}{c^4}\right)   \right) \text{ ,}
\end{equation}
\begin{equation}
Y(r) = \frac{1}{\left(1 - \frac{2 G M(r)}{c^2r}\right)} \left(-\frac{6}{r^2}   + \frac{4 \pi G}{ c^4}\left(5\epsilon + 9p + \frac{\epsilon + p}{(dp/d\epsilon)}\right) - \left(1 - \frac{2 G M(r)}{c^2r}\right)\left(\frac{d\nu}{dr}\right)^2   \right) \text{ ,}
\end{equation}
%where $\nu$ is the same radius dependent metric function~\ref{met_func}, where inside the body it depends on the structure as $d\nu/dr = -2(dp/dr)/(p + \epsilon)$. 
The dimensionless tidal deformability $\Lambda$ is written as $(2/3)k_2\left[(c^2/G)(R/M)\right]^5$. Using the above one can systematically find the mass, radius and $\Lambda$ of the neutron star.
\subsection{Formalism to compute the mass of the rotating neutron star}

The rotation deforms the neutron star: it flattens the star to some extent, depending upon its 
angular velocity $\Omega$. 
Formalism to compute the structure of the slowly rotating neutron stars is developed 
by Hartle and Thorne in their paper~\cite{1968ApJ...153..807H, 1967ApJ...150.1005H},
which we briefly review in this subsection.

The metric of a slowly rotating, stationary, and axially symmetric system may be defined as~\cite{1967ApJ...147..317H}:
\begin{equation}
ds^2 = -H^2dt^2 + Q^2dr^2 + r^2K^2\left[d\theta^2 + \sin^2\theta(d\phi - Ldt)^2\right] \text{ ,}
\end{equation}
\begin{center}
  \begin{figure}[t]
  \begin{subfigure}{.5\textwidth}
  \centering
  \includegraphics[width=1.0\linewidth]{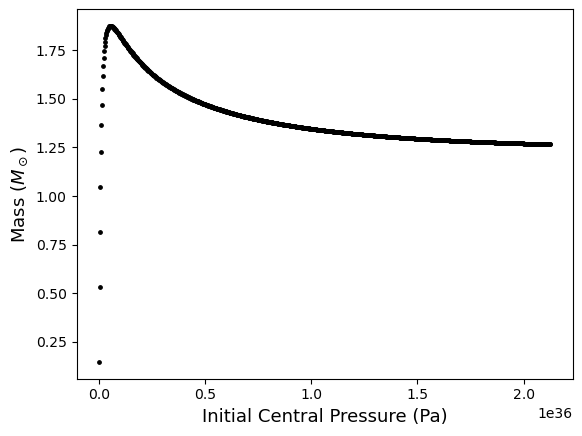}
  \caption{Mass-Pressure plot}
  \label{m_vs_p_b}
\end{subfigure}
\begin{subfigure}{.5\textwidth}
  \centering
  \includegraphics[width=1.0\linewidth]{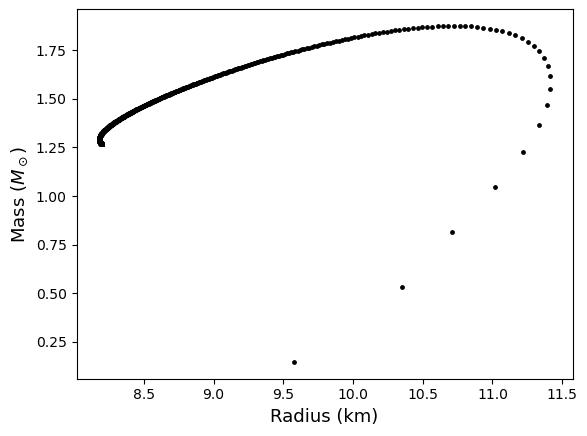}
  \caption{Mass-Radius plot}
  \label{m_vs_p_a}
\end{subfigure}%
\caption{(a) The Mass-Initial Central Pressure plots for calculating maximum mass and 
(b) The Mass-Radius plot of a Neutron Star}
\label{m_vs_p}
\end{figure}  
\end{center}
where $H, Q, K, L$ are functions of $r$ and $\theta$ alone. 
$L$ can be recognised as the angular velocity ($d\phi/dt$) acquired by an observer who falls from infinity to the point ($r,\theta$). The Einstein equations are then solved under this metric 
to derive various properties. 

Components of the Einstein equations at zeroth-order in $\Omega$ gives the TOV equation (Eq.~(\ref{TOV})).
We assume that the star's rotation is uniform, i.e., the angular velocity does not depend on
the position.
At first order in $\Omega$, the Einstein equations boil down to the following differential equation
for $\overline{\omega}(r,\theta) = \Omega - \omega(r,\theta)$ which is
the relative angular velocity of fluid with respect to the local inertial frame as:
%The first of the equations derived from this method allows us to form a differential equation on $\overline{\omega}(r,\theta) = \Omega - \omega(r,\theta)$, the relative angular velocity of fluid with respect to the local inertial frame as:
\begin{equation}
\frac{1}{r^4}\frac{d}{dr}\left(r^4j\frac{d\overline{\omega}}{dr}\right) + \frac{4}{r}\frac{dj}{dr}\overline{\omega} = 0 \text{ ,}
\label{eq_j}
\end{equation}
where $j = e^{-(\nu+\Phi)/2}$. By solving this equation, we can determine the metric up to first order in $\Omega$.

In order to compute the mass increase due to rotation, we need to consider
components of the Einstein equations at second order in $\Omega$. 
The mass increase $\delta M$ can be decomposed as \cite{1968ApJ...153..807H}
%We now need to calculate the mass increase in the neutron star following the rotation which can be written as:
\begin{equation}
\delta M = \frac{c^2m_0(R)}{G} + \frac{G J^2(\Omega)}{c^4 R^3} \text{ ,}
\end{equation}
where $J$ represents the angular momentum of star given by $c^2R^4(d\overline{\omega}/dr)/6G$ and $m_0(R)$ is obtained by solving the following coupled differential equations:
\begin{align}
&\frac{dm_0}{dr} = \left(\frac{4 \pi G r^2}{c^4}\left(\frac{d\epsilon}{dp}\right)(\epsilon + p)(p_0^*) + \frac{j^2 r^4}{12 c^2}\left(\frac{d\overline{w}}{dr}\right)^2 - \frac{2\overline{w}^2 r^3 j}{3 c^2}\left(\frac{dj}{dr}\right)\right),
\label{m0i} \\
&\frac{dp_0^*}{dr} = \left(-\frac{m_0\left(1 + \frac{8\pi G r^2 p}{c^4}\right)}{r^2\left(1 - \frac{2GM}{c^2r}\right)^2} - \frac{4 \pi G r (\epsilon +p)p_0^*}{c^4\left(1 - \frac{2GM}{c^2r}\right)}  +  \frac{j^2 r^3}{12 c^2\left(1 - \frac{2GM}{c^2r}\right)}\left(\frac{d\overline{w}}{dr}\right)^2 + \frac{1}{3c^2}\frac{d}{dr}\left(\frac{r^2j^2\overline{\omega}^2}{\left(1 - \frac{2GM}{c^2r}\right)}\right) \right) \text{ .}  
\label{p0i}
\end{align}
\begin{center}
\begin{figure}[h]
\centering
\includegraphics[width=0.65\linewidth]{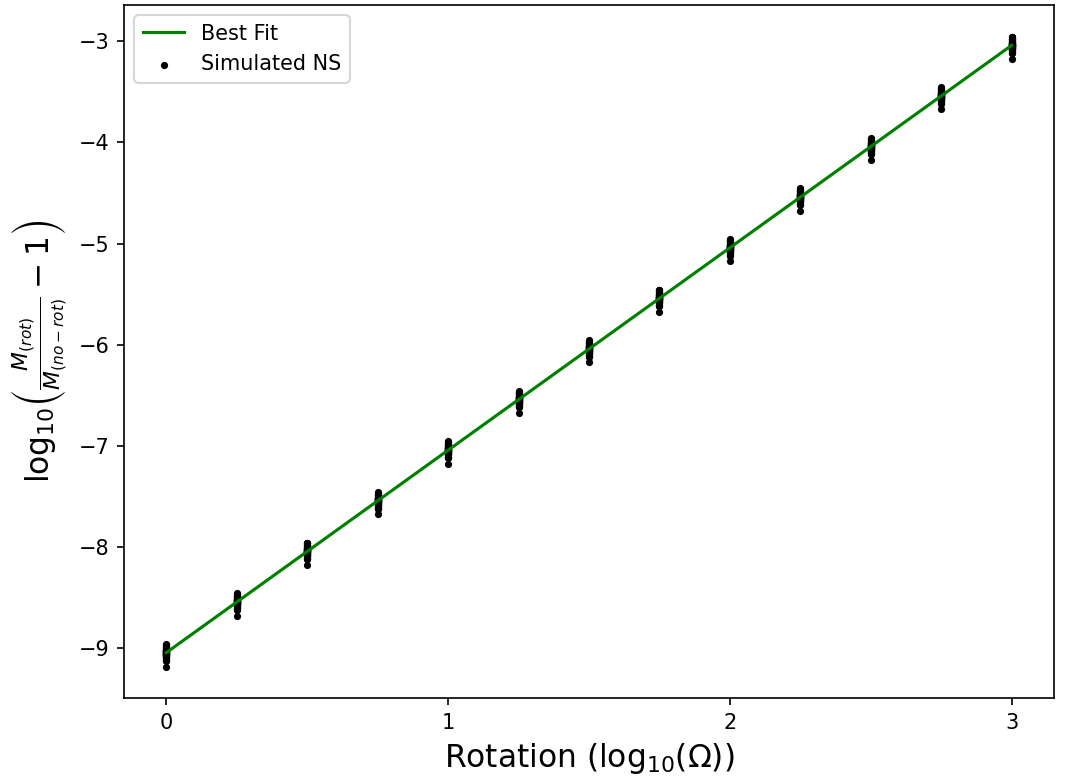}
\caption{The logarithm of the ratio of mass of rotating neutron stars to that of non-rotating one
as a function of the rotation velocity $\Omega$. Uniform rotation is assumed.}
\label{ratio_m}
\end{figure} 
\end{center}

\subsection{Mass-Radius plot and computing the Maximum Mass}
For our analysis here, we would need the maximum mass arising out of a single parameter set $\{n_0, B, K, S, L\}$. For this, we vary the initial starting pressure for a range of values, and see where the maximum appears. The effect of variation of initial central pressure on the mass of neutron stars has been shown in Fig.~\ref{m_vs_p_b}, from which the maximum mass is extracted upto the required accuracy. The Mass-radius plot is also depicted for the same set of parameters in Fig.~\ref{m_vs_p_a}.
\section{Results and Discussions}
\label{ReDi}
\subsection{Effect of rotation}
Firstly let us investigate the quantitative effect of rotation. 
To this end, we numerically solved the set of coupled equations~(\ref{eq_j})-~(\ref{p0i}) to obtain the ratio of mass of rotating neutron star to the one of non-rotating neutron star for various values
of $\Omega$. The result is plotted in Fig.~\ref{ratio_m}, in which at a single rotation value, we have composed 50 different random neutron stars simulations all markedly scattered very close to one point. The best fit line that passes through the mean of all these values has a slope of 2.007, which verifies the fact that $\delta M$ is proportional to $\Omega^2$~\cite{1968ApJ...153..807H}. In our calculations, we restricted ourselves to a maximum rotation frequency of $10^3$ Hz 
given the observations of $3~{\rm ms}$ and $40~{\rm ms}$ rotating massive neutron stars ~\cite{Antoniadis_2013,Demorest_2010}. 
%Given the observation of time period of rotation of massive neutron stars to be 3ms and 40ms~\cite{Antoniadis_2013,Demorest_2010}, we restrict our rotation to $10^{3}$ Hz. 
From Fig.~\ref{ratio_m}, we find that the relative mass increase is still ${\cal O}(0.1)\%$
even at the maximum rotation velocity we consider.
Since this level of tiny mass variation is irrelevant to our study, 
we neglect rotation in the following analysis.
%The histogram centers at a mean value of \textbf{put value}. 
\begin{center}
\begin{figure}[h]
\centering
\includegraphics[width=0.6\linewidth]{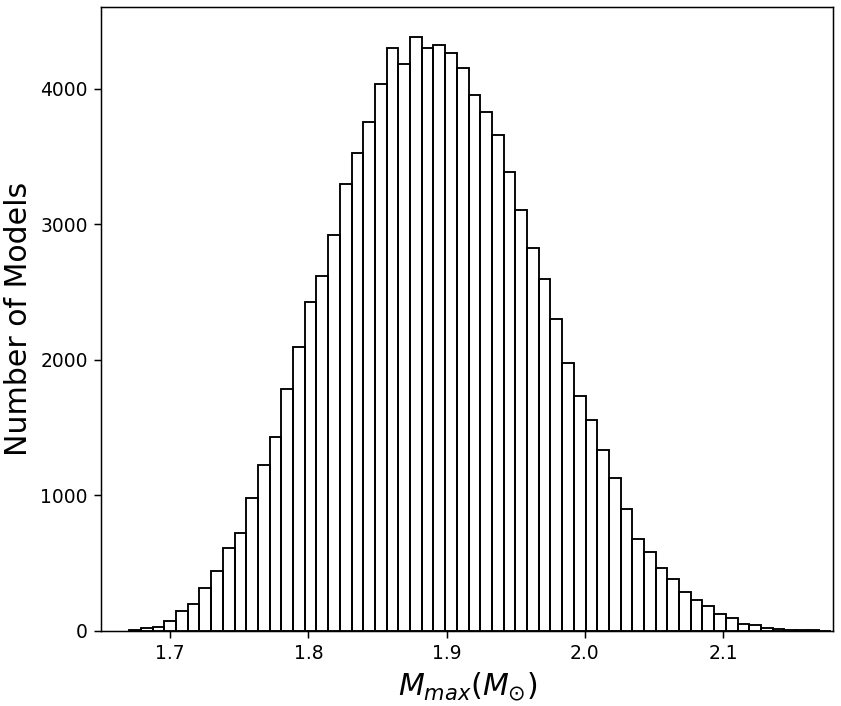}
\caption{The maximum mass histogram without any constraints}
\label{max_m_hist}
\end{figure} 
\end{center}
\vspace{-3em}
\subsection{Histogram of the maximum mass of the neutron star}
\label{POC}
As we have described in the previous section, we generate $10^5$ different EoS given by Eq.~(\ref{e-d}) 
by randomly selecting the parameters characterizing the EoS consistent with nuclear experiments.
For those EoS, we solve the TOV equation and obtain the curve in the mass-radius diagram 
like Fig.~\ref{m_vs_p_a}.
Fig.~\ref{max_m_hist} shows our result of the histogram of the maximum mass of the neutron star. The histogram centers around a mean value of $1.892~M_\odot$. We also notice that there are non-negligible fraction of EoS that achieve the maximum mass
exceeding $2~M_\odot$.
This may be ascribed to our choice of EoS (\ref{e-d}) whose pressure sharply increases beyond $2n_0$ to
realize the highest sound speed $v_s=1/\sqrt{3}c$.
We mention that our histogram does not exactly coincide with the one given in 
\cite{Bedaque:2014sqa} which also derived the histogram based on the same procedure as ours.
The origin of the discrepancy remains unsettled.
\begin{center}
\begin{figure}[h]
\centering
\includegraphics[width=0.7\linewidth]{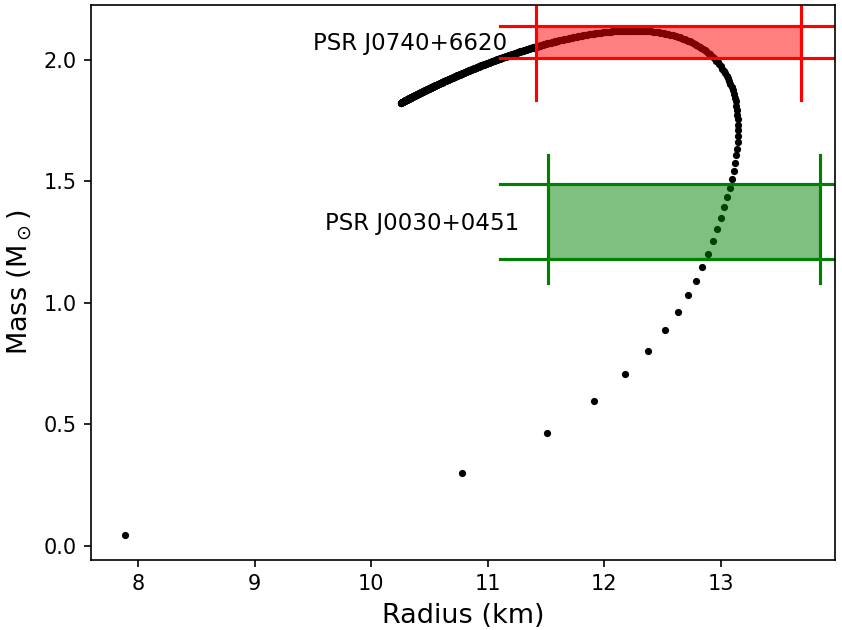}
\caption{The successful case of passing through the regions of uncertainty of both the neutron stars PSR J0030+0451 and PSR J0740+6620}
\label{suc_case}
\end{figure} 
\end{center}
We then proceed to place the following observational constraints on the mass-radius curve
for each EoS:
%The most intuitive constraint to place first is to check whether it lies in the range of a real Neutron Star. 
%We use the observational data from
i) NICER experiment which recently measured mass and radius of the two pulsars 
PSR J0030+0451 and PSR J0740+6620 as $1.34^{+0.15}_{-0.16} M_\odot$ and 
$2.072^{+0.067}_{-0.066} M_\odot$ for the mass respectively and 
$12.71^{+1.14}_{-1.19}$ km and $12.39^{+1.30}_{-0.98}$ km for the radius respectively~\cite{Riley:2019yda, Riley:2021pdl} \footnote{Although there are some other pulsars, such as PSR J1614-2230~\cite{Demorest_2010}, for which the mass ranges are known, we limit ourselves to the NICER data because it also contains the radius ranges, allowing for tighter constraints.}, ii)
tidal deformability $\Lambda$ constrained from the observation of the 
neutron star merger event GW170817 by the LIGO-Virgo Collaboration \cite{LIGOScientific:2017vwq}. 
We impose $\Lambda < 1400$ for the neutron star of mass 1.4 M$_\odot$(picked from the M-R diagram).
As an illustration, Fig.~\ref{suc_case}
shows the mass-radius relation of an EoS that passes the first condition i) and subsequently also the second ii).
%Fig.~\ref{suc_case} shows one such M-R diagram of the EoS starting from different pressures which passes the constraints put by both the observed Neutron Stars. We shall also place the constraint of tidal deformability $\Lambda$ taken from the neutron star merger event GW170817~\cite{LIGOScientific:2017vwq} for a mass of 1.4 M$_\odot$, which should be less than \textbf{put value}. 

\begin{center}
\begin{figure}[h]
\centering
\includegraphics[width=0.65\linewidth]{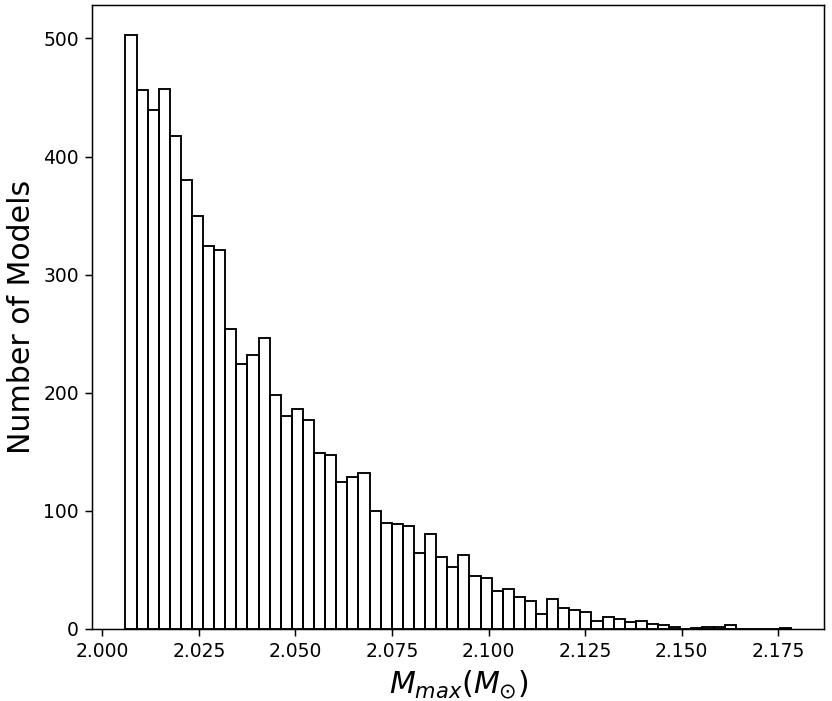}
\caption{The final histogram after applying the constraints from observed neutron masses and tidal deformability}
\label{fin-hist}
\end{figure} 
\end{center}

Fig.~\ref{fin-hist} is the histogram of the maximum mass of the neutron star
which has been obtained by applying the two observational constraints to Fig.~\ref{max_m_hist}.
First of all, we find that many models have been eliminated after the selection is applied, only 7.05 $\%$ models remained.
% Intriguingly, in the histogram, the maximum mass is at most $2.178~M_\odot$ which does not reach the
% lower end of the mass range of the recently detected pulsar ($M=2.35 \pm 0.17~M_\odot$) 
% \cite{Ecker:2022dlg}.
% Thus, the sound velocity bound $v_s \le \frac{1}{\sqrt{3}}c$ is in strong tension 
% with the existing pulsars.
Intriguingly, in the histogram, the maximum mass is at most $2.178~M_\odot$ which is very close to the lower end of the mass range of the recently detected pulsar ($M=2.35 \pm 0.17~M_\odot$)~\cite{Ecker:2022dlg}. Even if it is just one such set amongst $10^5$ parameter sets, we emphasize that the sound velocity bound is not ruled out robustly
because our sample is limited.
Even after we begin with $10^5$ different EoS, we end up with a few EoS that
achieve $M_{\rm max} \gtrsim 2.1~M_\odot$ after imposing the observational conditions.
In such a situation, different runs will give different tail in the histogram at large $M_{\rm max}$.
Unfortunately, the numerical computations with larger number of sample are not possible 
within reasonable time with our computers.
Thus, we leave the question as to whether the sound velocity bound is violated in nature an open issue.
Nevertheless, since the adopted EoS shows the sudden jump of its slope across $2n_0$, which
is highly artificial, more realistic EoS satisfying the sound velocity bound should be smoother.
This implies that the histogram based on such EoS shifts toward smaller $M_{\rm max}$, 
making the sound velocity bound more incompatible with observations. 
Thus, our analysis strong supports the idea that the sound velocity bound is violated
inside the heavy neutron stars observed in nature.
\vspace{-3em}

\section{Conclusion}
Observations of the massive neutron stars indicate that the pressure 
of nuclear matter rises rapidly as the energy density is increased at extremely large density. 
In particular, it has been suggested in the literature that the sound velocity
bound $v_s \le 1/\sqrt{3} c$ is violated at such high density.
In this paper, we have studied to what extent the current measurements of  
mass, radius, and tidal deformability of neutron stars reinforces the violation
of the sound velocity bound.
In order to minimize the impact of the uncertainties of the equation of state of
nuclear matter and derive a conservative consequence, 
we employed the equation of state that is parameterized by a set of parameters 
below twice the saturation density and saturates the sound velocity bound above that density
and obtained the mass-radius relation for $10^5$ different sets of parameters.
We found majority of the models are eliminated by the incompatibility with the observations.
Especially, the recently detected massive pulsar ($2.35\pm 0.17 M_\odot$) is hardly realized
by our simulations.
Our study strongly supports the violation of the sound velocity bound.

\section*{Acknowledgements}
T.S.\ was supported by the MEXT Grant-in-Aid for Scientific Research on Innovative Areas No.\ 17H06359, No.\ 19K03864, and No.\ 21H05453.

\bibliography{ref}
\end{document}